\documentclass[a4paper,11pt,final]{scrartcl}
\usepackage[utf8]{inputenc}
\usepackage{ifpdf}
\usepackage{latexsym}
\usepackage{exscale}
\usepackage{amsfonts}
\usepackage{eucal}
\usepackage{cmmib57}
\usepackage[centertags]{amsmath}
\usepackage{amssymb}
\usepackage{amsbsy}
\usepackage{amstext}
\usepackage{amsthm,thmtools}
\usepackage{xcolor}
\usepackage{framed}
\usepackage{tocbibind}
\usepackage{tocloft}
\usepackage{etoolbox}
\usepackage[ruled,vlined,algosection]{algorithm2e}
\usepackage{delarray}
\usepackage[automark]{scrpage2}
\usepackage{times}
\usepackage[toc,page]{appendix}
\usepackage{textcomp}
\usepackage{mflogo}
\usepackage{here}
\usepackage[round]{natbib}
\usepackage{here}
\usepackage{longtable}
\usepackage{threeparttablex}
\usepackage{multirow}
\usepackage{multicol}
\usepackage{caption}
\usepackage{lscape}
\usepackage{rotating,capt-of}
\usepackage[margin=15mm]{geometry}
\usepackage{pgf}
\usepackage{tikz}
\usetikzlibrary{arrows,automata,shapes}
\usepackage{verbatim}
\newcounter{treeline}

\ifvtexpdf\pdftrue\fi
\ifpdf
\usepackage{graphicx}
\usepackage[pdftex]{thumbpdf}
\usepackage{nameref}
\usepackage[pdftex,colorlinks]{hyperref}
\hypersetup{
           breaklinks=true,
           pdfpagemode=UseOutlines,
           colorlinks=true,
           linkcolor=blue,
           citecolor=blue,
           bookmarksopen=true,
           pdftitle={Finding the nucleoli of large cooperative games: A Disproof with Counter-Example},
           pdfauthor={Holger Ingmar Meinhardt},
           pdfsubject={Tu Games, Propositional Logic},
           pdfkeywords={Axiomatization, Propositional Logic, Material Implication, Indirect Proof, Tu Games},
}

\else
\usepackage{epsfig}
\usepackage[ps2pdf]{thumbpdf}
\usepackage{nameref}
\usepackage[ps2pdf]{hyperref}

\hypersetup{
           breaklinks=true,
           pdfpagemode=UseOutlines,
           colorlinks=true,
           linkcolor=blue,
           citecolor=blue,
           bookmarksopen=true,
           pdftitle={Finding the nucleoli of large cooperative games: A Disproof with Counter-Example},
           pdfauthor={Holger Ingmar Meinhardt},
           pdfsubject={Matlab Game Theory Toolbox, Tu Games},
           pdfkeywords={Axiomatization, Propositional Logic, Material Implication, Indirect Proof, Tu Games},
}
\fi

\addtocounter{MaxMatrixCols}{20}

\makeatletter
\renewcommand{\section}{\@startsection
  {section}%
  {1}%
  {0em}%
  {-\baselineskip}%
  {0.5\baselineskip}%
  {\centering\normalfont\Large\scshape\mdseries}}%
\makeatother

\makeatletter
\renewcommand{\subsection}{\@startsection
  {subsection}%
  {2}%
  {0em}%
  {-\baselineskip}%
  {0.5\baselineskip}%
  {\normalfont\large\scshape\mdseries}}%
\makeatother

\makeatletter
\renewcommand*\env@matrix[1][c]{\hskip -\arraycolsep
  \let\@ifnextchar\new@ifnextchar
  \array{*\c@MaxMatrixCols #1}}
\makeatother

\makeatletter

\newenvironment{theopargself*}
    {\def\@spopargbegintheorem##1##2##3##4##5{\trivlist
         \item[\hskip\labelsep{##4##1\ ##2}]{\hspace*{-\labelsep}##4##3\@thmcounterend}##5}
     \def\@Opargbegintheorem##1##2##3##4{##4\trivlist
         \item[\hskip\labelsep{##3##1}]{\hspace*{-\labelsep}##3##2\@thmcounterend}}}{}
\def \@floatboxreset {%
        \reset@font
        \small
        \@setnobreak
        \@setminipage
}
\def\figure{\@float{figure}}
\@namedef{figure*}{\@dblfloat{figure}}
\def\table{\@float{table}}
\@namedef{table*}{\@dblfloat{table}}
\def\fps@figure{htbp}
\def\fps@table{htbp}

\makeatother

\makeatletter
\@addtoreset{equation}{section}
\makeatother
\makeatletter
\@addtoreset{table}{section}
\makeatother

\theoremstyle{plain}

\newtheoremstyle{break}
  {9pt}
  {9pt}
  {\itshape}
  {}
  {\bfseries}
  {.}
  {\newline}
  {}

\newtheoremstyle{break1}
  {9pt}
  {9pt}
  {\rmfamily}
  {}
  {\scshape}
  {.}
  {\newline}
  {}

\theoremstyle{break}

\newtheoremstyle{note}
  {3pt}
  {3pt}
  {}
  {}
  {\itshape}
  {:}
  {.5em}
  {\newline}  
  {}

\theoremstyle{note}

\theoremstyle{definition}
\newtheorem{example}{Example}[section]

\theoremstyle{break1}

\begin{document}
\bibliographystyle{plainnat} 
\pdfbookmark[0]{Finding the nucleoli of large cooperative games: A Disproof with Counter-Example}{tit}

\title{{Finding the Nucleoli of Large Cooperative Games: A Disproof with Counter-Example} 
}
\author{{\bfseries Holger I. MEINHARDT}~\thanks{Holger I. Meinhardt, Institute of Operations Research, Karlsruhe Institute of Technology (KIT), Englerstr. 11, Building: 11.40, D-76128 Karlsruhe. E-mail: \href{mailto:Holger.Meinhardt@wiwi.uni-karlsruhe.de}{Holger.Meinhardt@wiwi.uni-karlsruhe.de}} 
}
\maketitle

\begin{abstract}
\citet{Nguyen:16} claimed that they have found a method to compute the nucleoli of games with more than $50$ players using nested linear programs (LP). Unfortunately, this claim is false. They incorrectly applied the indirect proof by ``{\itshape $A \land$ not $B$} implies {\itshape $A \land$ not $A$}'' to conclude that ``{\itshape if $A$ then $B$}''is valid. In fact, they prove that a truth implies a falsehood. As established by~\citet{mei:15}, this is a wrong statement. Therefore, instead of giving a proof of their main Theorem 4b, they give a disproof. It comes as no surprise to us that the flow game example presented by these authors to support their arguments is obviously a counter-example of their algorithm. We show that the computed solution by this algorithm is neither the nucleolus nor a core element of the flow game. Moreover, the stopping criterion of all proposed methods is wrong, since it does not satisfy one of Kohlberg's properties (cf.~\citet{kohlb:71}). As a consequence, none of these algorithms is robust.\\


\end{abstract}


\thispagestyle{empty}
\pagebreak

\pagestyle{scrheadings}  \ihead{\empty} \chead{Finding the nucleoli: A Disproof with Counter-Example} \ohead{\empty}

\section{Introduction}
\label{sec:intod}
\citet{Nguyen:16} claimed that they have invented a method for the extremely challenging task of computing the nucleoli of games with more than $50$ players using nested linear programs (LP). Unfortunately, their claim is false. They incorrectly applied the indirect proof for Theorem 4b by assuming the premise {\itshape $A \land \neg B$} to derive a falsum $\bot$, namely {\itshape $A \land \neg A$}, to conclude that {\itshape $\neg B \Rightarrow \neg A$} holds, and therefore, they infer {\itshape $A \Rightarrow B$} is a valid statement. In fact, they prove that a truth implies a falsehood, which is a wrong assertion implying that they gave for their main theorem a disproof. For more details about the logical background, we refer the inclined reader to~\citet{mei:15}.

Notice that the statement {\itshape if $A \Rightarrow B$} and its contrapositive {\itshape if $\neg B \Rightarrow \neg A$} are logically equivalent statements, which are also equivalent to the disjunction $\neg A \lor B$. To prove the implication {\itshape if $A \Rightarrow B$}, we can focus on the opposite {\itshape $\neg (A \Rightarrow B)\equiv \neg (\neg A \lor B) \equiv$ $(A \land \neg B)$} in order to get from {\itshape if $A \Rightarrow B$} the logical equivalent implication {\itshape if $A \land \neg B \Rightarrow B \land \neg B$}. This imposes a proof by contradiction, since $B \land \neg B$ is a falsum $\bot$. However, if the starting point is a proof by contraposition, i.e., $\neg B \Rightarrow \neg A$, we obtain the following equivalent statement $A \land \neg B \Rightarrow A \land \neg A$. It should be evident that this also imposes a proof by contradiction.    

To observe that~\citeauthor{Nguyen:16} have incorrectly applied the indirect proof, we rewrite the equivalent relationship of $(A \land \neg B \Rightarrow A \land \neg A) \equiv (A \Rightarrow B)$ more concisely as $(\phi \Rightarrow \bot) \Leftrightarrow \neg \phi$. If we now suppose as the authors that the premise $\phi$ is true, we know immediately that $\neg \phi$ must be false, and therefore the statement $\phi \Rightarrow \bot$ must be false either, due to the above equivalence. But then we can only infer that $A \Rightarrow B$ is a falsehood, i.e., $A \not\Rightarrow B$. In contrast, to get a valid statement from $(\phi \Rightarrow \bot) \Leftrightarrow \neg \phi$ one has to set the premise $\phi$ to false to know that $\neg \phi$ is true, but due to the above equivalence $\phi \Rightarrow \bot$ must be true as well, and from this we infer that $A \Rightarrow B$ is a truth.   

However, it is an impermissible conduct to assume that $A \Rightarrow B$ is false, i.e., $A \land \neg B$ holds in order to derive a contradiction, say $A \land \neg A$, to finally deduce from this contradiction that $A \land \neg B$ is false, and that one has therefore proved $A \Rightarrow B$ by the logical equivalence of $A \land \neg B \Rightarrow A \land \neg A$ and $A \Rightarrow B$. Doing so, means that we always get the desired result, and we could prove perverted results (see, for instance, Example~\ref{exp:elm}). Of course, this is a fallacy and one has disproved oneself, since one gets that $A \land \neg B \Rightarrow A \land \neg A$ is a falsehood confirming that $A \Rightarrow B$ is false as well. Obviously, this kind of arguing is a circular reasoning (circulus in probando). Unfortunately, this is exactly the line of argument used by~\citeauthor{Nguyen:16}. They have shown in their proof for Theorem 4b the exact opposite of what had been intended to prove. 

The remaining part of this note is organized as follows. In Section~\ref{sec:dispr} we present in a first step a short reminder of formal logic, before we discuss in more details the purported proof of Theorem 4b. Establishing by the above logical arguments that the main result of their article is flawed. After having clarified this issue, we turn to the flow game example of~\citet{Nguyen:16} to show that the proposed solution by the authors is even not in the core of the game, and that it cannot be from this point of view the nucleolus of the game. We close this note by some final remarks in Section~\ref{sec:disrem}.

\section{The Disproof of Theorem 4b with Counter-Example}
\label{sec:dispr}

In the sequel, we rely on the same notation and definitions as they can be found in~\citet{Nguyen:16}. However, for the understanding of our arguments these are not really needed. Only some basics from formal logic is needed, which we start to refresh in the next paragraph.

For presenting a reminder of propositional logic, we introduce two truth tables. A logical statement/proposition is formed by the symbols $A$ or $B$, which means that a statement $A$ is true or false. However, the inversion is formed by the negation of a proposition by using the logical term ``not'' denoted by $\neg$. If $A$ is a proposition, then $\neg A$ is the negation of $A$ verbalized as ``not $A$'' or ``$A$ is false''. The effect of negation, conjunction, disjunction, and implication on the truth values of logical statements is summarized by a so-called truth table. In this table, the capital letter {\bfseries T} indicates a true proposition and {\bfseries F} indicates that it is false. 

\begin{center}
\begin{tabular}{c c c c c c c c c c c}
\hline
$A$ & $B$ & $\neg B$ & $A \Rightarrow B$ & $ \neg (A \Rightarrow B)$ & $A \Leftarrow B$ & $A \Leftrightarrow B$ & $A \lor \neg B$ & $A \land B $ &  $A \lor B $\\
\hline
F & F & T & T & F & T & T & T & F & F \\
F & T & F & T & F & F & F & F & F & T \\
T & F & T & F & T & T & F & T & F & T \\
T & T & F & T & F & T & T & T & T & T \\ 
\end{tabular}

\begin{tabular}{c c c c |c c ||c c| c c}
\hline
$A$ & $B$ &  $\neg A$ & $\neg B$ & $\neg A \Rightarrow \neg B$ & $A \lor \neg B$ & $\neg A \Leftarrow \neg B$ & $\neg A \lor B$  & $A \land \neg B$ & $\neg A \Leftrightarrow \neg B$ \\
\hline
F & F & T & T & T & T & T & T & F & T   \\
F & T & T & F & F & F & T & T & F & F   \\
T & F & F & T & T & T & F & F & T & F   \\
T & T & F & F & T & T & T & T & F & T   \\\hline
\end{tabular}
\end{center}
Two statements are indicated as logically equivalent through the symbol $\equiv$. For instance, by the truth table we realize that the two statements $\neg A \Leftarrow \neg B$ and $\neg A \lor B$ are logically equivalent, which is formally expressed by $(\neg A \Leftarrow \neg B) \equiv (\neg A \lor B)$. A falsum $\bot$ is, for instance, the conjunction $A \land \neg A$ whereas a verum (tautology) $\top$ can be expressed, for instance, by the disjunction $\neg A \lor A$. For more details see~\citet{mei:15}. 

To see that from a false conclusion a false implication follows, can be observed from an example taken from an elementary course in mathematics, which we have reused from~\citet{mei:15}. 

\begin{example}
 \label{exp:elm}
Let $m$ denote an arbitrary number, and let us ``{\itshape prove}'' the wrong implication that

if $m^2$ is even ($A$), then $m$ is odd ($B$),

\noindent while running a purported proof. In a first step, we assume that $A \land \neg B$ is valid. For this purpose, we suppose that $m$ is even ($\neg B$) s.t. $m=2\;k$ for some integer $k$, and assume that $m^2$ is even too ($A$ is true), i.e., $m^2=2\;q$ for some integer $q$, then we get that $m^2=(2\;k)^2=4\;k^2 = 2\;q$. This implies $k=\pm\,\sqrt(q/2)$, which is the {\itshape desired contradiction}. We conclude that $m$ is odd ($B$). Hence, a valid premise $A \land \neg B$ implies something wrong ($B \land \neg B$). In the literature, it is a commonly held believe that this is a true proposition. Assuming this, one would conclude that $A \land \neg B$ is wrong, then the negation of this expression, i.e., $\neg A \lor B$ is true. From which one would deduce that $A \Rightarrow B$ is a valid statement. This is certainly a fallacy, one incorrectly applied $(\phi \Rightarrow \bot) \Leftrightarrow \neg \phi$. However, it should be obvious by the discussion from the introduction that this gives in fact a disproof of $A \Rightarrow B$, thus we have $A \not\Rightarrow B$.
\end{example} 

We quote the main Theorem 4 from~\citet{Nguyen:16} and discuss their proof in order to observe how deficient these authors have applied the indirect proof. We cite only the essential parts and conclusions of the authors, and set their wrong arguments in italic and highlighted them by a red coloring.    

\begin{quote}
\begin{labeling}[:]{Theorem}
\item[\bfseries{Theorem 4.~(\citet[pp.~1086-87]{Nguyen:16})}]
\itshape{Let $(\mathbf{x},\widehat{\epsilon^{*}_{k}})$ be any optimal solution of $\widehat{\widehat{\mathbf{LP}}}_{k}$ with the tight set $\widehat{\tau}(\mathbf{x})$. Then solving {\bfseries FBOS2} will lead to one of two cases:}
\begin{itemize}
\item[(a)] If $\mathbf{x}$ is not an optimal solution (\ldots)  smaller tight set, i.e., $\arrowvert \widehat{\tau_{k}}(\mathbf{y^{*}}) \arrowvert < \arrowvert \widehat{\tau_{k}}(\mathbf{x^{*}}) \arrowvert$.
\item[(b)] If $\mathbf{x}$ is an optimal solution of {\bfseries FBOS2}, then $\mathbf{x}$ is also an optimal solution of $\widehat{\widehat{\mathbf{LP}}}_{k}$ with the minimal tight set, i.e., $\widehat{\tau}_{k}(\mathbf{x})=\widehat{\tau^{*}_{k}}$.
\end{itemize} 
\end{labeling}
\begin{proof}
Let $R^{*}_{k}(\mathbf{x})=\{\mathbf{z}^{(1)},\mathbf{z}^{(2)},\ldots,\mathbf{z}^{(r)}\}$.
\begin{itemize}
\item[(a)] Notice that $\mathbf{x},\widehat{\mathbf{x}}^{*}_{k}$ are feasible (\ldots) compared to $\mathbf{x}$ in $\widehat{\mathbf{LP}}_{k}$.
\item[(b)] {\itshape \color{red} Suppose $\mathbf{x}$ is an optimal solution of {\bfseries FBOS2}}. We will prove $\mathbf{x}$ is also an optimal solution of $\widehat{\widehat{\mathbf{LP}}}_{k}$ with the minimal tight set. {\itshape \color{red} Suppose, as a contradiction, that there exists another optimal solution $\mathbf{y}^{*}$ of $\widehat{\widehat{\mathbf{LP}}}_{k}$ with $\arrowvert \widehat{\tau_{k}}(\mathbf{y^{*}}) \arrowvert < \arrowvert \widehat{\tau_{k}}(\mathbf{x^{*}}) \arrowvert$}. This means there exists at least (\ldots). We can choose $\alpha_{0}>0$ and small enough such that $\alpha_{i} \ge 0, \;\forall\; i$. Thus,
  \begin{equation*}
    \mathbf{c}^{t}(\mathbf{x}-\mathbf{y}^{*}) = \underbrace{\sum^{r}_{i=1}\,\alpha_{i}(\mathbf{z^{(i)}})^{t}(\mathbf{x}-\mathbf{y}^{*})}_{\le 0} + \underbrace{\alpha_{0}(\mathbf{z^{(0)}})^{t}(\mathbf{x}-\mathbf{y}^{*})}_{<0}<0.
  \end{equation*}
{\itshape \color{red} This means $\mathbf{x}$ is not an optimal solution of {\bfseries FBOS2} and so there is a contradiction}. The proof of the theorem is complete. (\citet[p.~1087]{Nguyen:16})
\end{itemize}
\end{proof}
\end{quote}
We give now the reasons why Theorem 4b cannot be correct. For doing so, let us restate the assertion of Theorem 4b by it essential parts:

\noindent If $\mathbf{x}$ is an optimal solution of {\bfseries FBOS2} ($A$ is true), then $\mathbf{x}$ is also an optimal solution of $\widehat{\widehat{\mathbf{LP}}}_{k}$ ($B$ is true).

They started by assuming that $\mathbf{x}$ is an optimal solution of {\bfseries FBOS2}, hence $(A)$ is true. In the next step, they assume that there exists an optimal solution $\mathbf{y}^{*}$ of $\widehat{\widehat{\mathbf{LP}}}_{k}$. Thus, they set $(\neg B)$ by supposition. Through imposition of the assumptions the authors have set the premise $A \land \neg B$ as valid. The authors proceed to derive the contradiction $\mathbf{c}^{t}(\mathbf{x}-\mathbf{y}^{*})<0$.  From this outcome, they conclude that $\mathbf{x}$ is not an optimal solution of {\bfseries FBOS2} in contrast to their initial assumption. They get that $\neg A$ must be satisfied. This is their {\itshape desired contradiction} and they conclude that $A \Rightarrow B$ must hold. Of course, this is a fallacy. 

By the consideration from above, we realize that~\citeauthor{Nguyen:16} have shown that a valid premise $A \land \neg B$ implies the falsum $A \land \neg A$. The authors have established that a truth implies a falsehood, which is a wrong statement. One incorrectly applied $(\phi \Rightarrow \bot) \Leftrightarrow \neg \phi$. This is due to the fact that they have set $\phi:=(A \land \neg B)$ as valid, from which follows that $(A \Rightarrow B)=:\neg \phi$ must be false either, and therefore $(\phi \Rightarrow \bot)$ must be false as well by equivalence of $(\phi \Rightarrow \bot) \Leftrightarrow \neg \phi$. By these arguments, we observe that $(\phi \Rightarrow \bot)$ cannot be a truth in contrast to their implicit assumption. Hence, we have $A \not\Rightarrow B$, and the authors have imposed a circular reasoning or a circulus in probando.

Remember that the implications $A \land \neg B \Rightarrow A \land \neg A$ or $A \land \neg B \Rightarrow B \land \neg B$ are logically equivalent to $A \Rightarrow B$. Hence, if one has shown that such an implication or every other implication that should be equivalent to $A \Rightarrow B$ produces a wrong proposition, one has to conclude that $A \Rightarrow B$ must be invalid too. In this case, one cannot deduce that $A \land \neg B$ is false, this is due to that $A \land \neg B$ was assumed to be valid. Applying then that $A \land \neg B$ is false in order to infer from this, that its negation $A \lor \neg B$ as well as the implication $A \Rightarrow B$ must be valid, is, by the above consideration, a fallacy.  

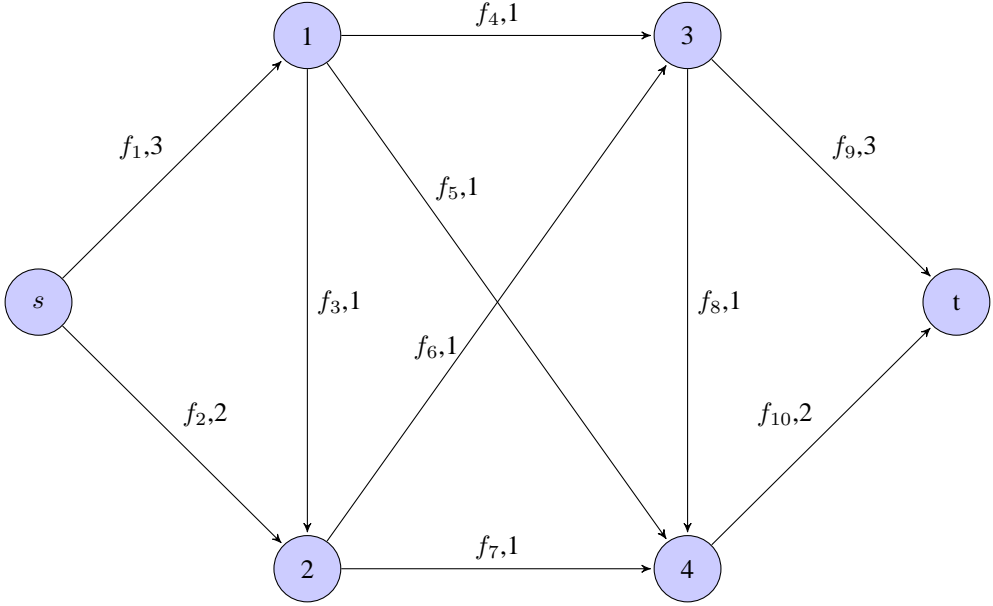
\begin{figure}[H]
\begin{minipage}[t]{14cm}
\begin{center}
\begin{tikzpicture}
  [->,>=stealth',shorten >=1pt,auto=left,node distance=5.0cm,scale=.8,every state/.style={circle,fill=blue!20}]
  \node[state] (n0)           {$s$};
  \node[state] (n1) [above right of=n0] {1};
  \node[state] (n2) [below right of=n0] {2};
  \node[state] (n3) [right of=n1]  {3};
  \node[state] (n4) [right of=n2]  {4};
  \node[state] (n5) [below right of=n3] {t};

  \path (n0) edge              node {$f_{1}$,3} (n1)
             edge              node {$f_{2}$,2} (n2)
        (n1) edge              node {$f_{3}$,1} (n2)
             edge              node {$f_{4}$,1} (n3)
            edge               node[xshift=-2.7em,yshift=3.5em] {$f_{5}$,1} (n4) 
        (n2) edge              node[xshift=-1.1em,yshift=-2.5em] {$f_{6}$,1} (n3)
             edge              node {$f_{7}$,1} (n4)
        (n3) edge              node {$f_{8}$,1} (n4)
             edge              node {$f_{9}$,3} (n5)
        (n4) edge              node {$f_{10}$,2} (n5);

\end{tikzpicture}
\caption{Flow Problem of 10 persons and 6 nodes}
\label{fig:flow01}
\end{center}
\end{minipage}
\end{figure}

Now, let us turn to the flow game example from which~\citeauthor{Nguyen:16} want to demonstrate the strength of their Algorithm 3 to successfully find the nucleolus. They introduce a flow game problem with $6$-nodes and $10$-persons. The persons are numbered at the edges in accordance with $f_{i}$ in the Figure~\ref{fig:flow01}, which we have reproduced from their article. The capacities of the edges are $c_{1}=c_{9}=3$, $c_{i}=1,\,\forall i \in \{3,\ldots,8\} $ and $c_{2}=c_{10}=2$. Theses values can be found at the second position at the labeling indicating in the first place the owner of an edge. 

The optimal solution found after some iterations by their Algorithm 3 is given by the authors as $\mathbf{x}^{*}_{2}=\{1,0.2,0,0.2,0.4,0.4,0.6,0,1,0.2\}$ (cf.~\citet[pp.~1088-89]{Nguyen:16}). In addition, they claim that the corresponding minimal set is tight, from which they conclude that the solution must be the nucleolus of the coalitional game arising from the flow problem of Figure~\ref{fig:flow01}. 

In order to reproduce their solution related to the nucleolus, we have written a small MATLAB program to compute the corresponding transferable utility game from the above flow problem.\footnote{This MATLAB program computes from an arbitrary flow problem the associated TU game, which we can make available upon request.} For a cross-check that our flow game is correctly specified, we relied on the minimal cuts. From Figure~\ref{fig:flow01}, we observe that there are only 2 minimal cuts: namely $f_{4},f_{5},f_{6},f_{7}$ and $f_{4},f_{6},f_{10}$ having both a flow of 4. A removal of these cuts from the graph prevented a flow of positive magnitude from the source $c$ to the sink $t$, which we observed. Moreover, the capacities of these minimal cuts are core elements.  

From this game, since flow games are totally balanced, we can check whether the proposed solution is at least a core element before we go a step further to verify one of Kohlberg's properties (cf.~\citet{kohlb:71}). For this purpose, we look on blocking coalitions. By Figure~\ref{fig:flow01}, we observe that a blocking coalition is $\{1,4,5,8,10\}$.\footnote{In total, we observe 10 blocking coalitions.} This is due to that the maximal flow the sub-coalition $\{1,4,5,8,10\}$ can generate is $2$, and from this observation, we see that $v(\{1,4,5,8,10\}) - \mathbf{x}^{*}_{2}(\{1,4,5,8,10\}) = 2 - 1.8 = 0.2 >0$ must hold. Therefore, the proposed solution $\mathbf{x}^{*}_{2}$ is not a core element, but then it cannot be the nucleolus. In accordance with the fact that the flow game is zero-monotonic, we can even conclude that the solution is neither the pre-nucleolus nor a pre-kernel nor a kernel point. Hence, to conclude solely from the full rank condition of $\widehat{\mathcal{H}}^{*}_{3}$ that one has found the nucleolus is a fallacy, one also has to check that the induced collection of coalitions from $\widehat{\mathcal{H}}^{*}_{3}$ is balanced (cf.~\citet{kohlb:71}). This means, in addition, that the stopping criterion of all Algorithms is false, and the proof of Theorem 2 is not complete. Rather than $\mathbf{x}^{*}_{2}$, the nucleolus $\nu$ of the game is $\{11/15,1/5,0,1/3,1/5,3/5,1/3,0,8/15,16/15\}$.\footnote{Computed solution confirmed as the nucleolus of the flow game from the authors by private conversation.} This solution satisfies efficiency $v(N)=x(N)=4$ as well as Kohlberg's property II, it is in addition an element of the kernel as well as a core element, hence exemplarily we have $v(\{1,4,5,8,10\}) - \nu(\{1,4,5,8,10\}) = 2 - 7/3 = -1/3 <0$. To check it for all coalitions, we refer the reader to~\citet{mei:15a}.\footnote{Apart from checking that an imputation belongs to the core, one can also verify by this MATLAB toolbox, for instance, if Kohlberg's property II or the (pre-)kernel properties are satisfied. The documentation of the toolbox is given by~\citet{mei:11a} and ships with the toolbox.} We conclude that the proposed solution of~\citeauthor{Nguyen:16} is not the nucleolus of their flow game, and that Algorithm 3 neither finds the nucleoli of small games nor that it can be used to calculate nucleoli for large games with more than $50$ players. 

Finally, in their computer experiment involving large weighted voting and coalitional skill games, the authors have not presented any evidence that the computed solutions satisfy one of Kohlberg's properties. This provides further evidence that Algorithm 3 is not applicable in the computation of the nucleoli of cooperative games.     

\section{Concluding Remarks}
\label{sec:disrem}
We have found in the article of~\citet{Nguyen:16} severe deficiencies so that the reported results become invalid. First of all, the proof of their main Theorem 4b is logically flawed, because of an incorrect application of the indirect proof. Secondly, the presented example that should demonstrate the strength of their algorithm is in fact a counter-example. Finally, the stopping criterion of all proposed methods does not take into account one of Kohlberg's properties. Therefore, the authors have not imposed the correct stopping criterion. All of this invalidates the results of the article.

\pagestyle{scrheadings} \chead{\empty}  
\footnotesize
\bibliography{findnuc}

\end{document}